\documentclass{emulateapj}

\shorttitle{Anomalous emission in the Pleiades}
\shortauthors{G\'enova-Santos et al.}

\begin{document}

\title{Detection of anomalous microwave emission in the Pleiades reflection nebula with WMAP and the COSMOSOMAS experiment}
\author{R.~G\'enova-Santos\altaffilmark{1,2}, R.~Rebolo\altaffilmark{1,2,3}, J.~A.~Rubi\~no-Mart\'{\i}n\altaffilmark{1,2}, 
C.~H.~L\'opez-Caraballo\altaffilmark{1,2}, S.~R.~Hildebrandt\altaffilmark{1,4}}
\affil{Instituto de Astrof\'{\i}sica de Canarias, C/ V\'{\i}a L\'actea s/n, E-38200, La Laguna, Tenerife, Spain.}
\affil{Departamento de Astrof\'{\i}sica, Universidad de La Laguna (ULL), E-38206 La Laguna, Tenerife, Spain.}
\affil{Consejo Superior de Investigaciones Cient\'{\i}ficas, Spain.}
\affil{California Institute of Technology, 1200 E California Boulevard 91125, Pasadena, California, USA.}

\begin{abstract}
We present evidence for anomalous microwave emission (AME) in the Pleiades reflection nebula, using data from the 
seven-year release of the {\it Wilkinson Microwave Anisotropy Probe} (WMAP) and from the COSMOSOMAS experiment.
The flux integrated in a 1$\degr$ radius around R.A.=56.24$\degr$, Dec.=23.78$\degr$ (J2000) is $2.15\pm 0.12$~Jy 
at 22.8~GHz, where AME is dominant. COSMOSOMAS data show no significant emission, but allow to set upper limits of 
0.94 and 1.58~Jy (99.7\% C.L.) respectively at 10.9 and 14.7~GHz, 
which are crucial to pin down the AME spectrum at these frequencies,  
and to discard any other emission mechanisms which could have an important contribution to the signal detected at 22.8~GHz. 
We estimate the expected level of free-free emission from an extinction-corrected H$\alpha$ template, while the thermal 
dust emission is characterized from infrared DIRBE data and extrapolated to microwave frequencies. When we deduct the contribution 
from these two components at 22.8~GHz the residual flux, associated with AME, is $2.12\pm 0.12$~Jy (17.7$\sigma$).
The spectral energy distribution from 10 to 60~GHz can be accurately fitted with a model of electric dipole 
emission from small spinning dust grains distributed in two separated phases of molecular and atomic gas, respectively.

The dust emissivity, calculated by correlating the 22.8~GHz data with 100~$\mu$m data, is found to be 
$4.36\pm 0.17$~$\mu$K/(MJy~sr$^{-1}$), 
a value considerably lower than in typical AME clouds, which present 
emissivities of $\sim 20$~$\mu$K/(MJy~sr$^{-1}$), although higher than the $0.2$~$\mu$K/(MJy~sr$^{-1}$) of the translucent cloud 
LDN~1780, where AME has recently been claimed.
The physical properties of the Pleiades nebula, in particular 
its low extinction $A_{\rm V}\sim 0.4$, indicate that this is indeed a much less opaque object than those were 
AME has usually been studied. This fact, together with the broad knowledge of the stellar content of this region, 
provides an excellent testbed for AME characterization in physical conditions different from those generally explored up to now.
\end{abstract}


\section{Introduction}{\label{sec:intro}}
Several microwave experiments, like {\it COBE}-DMR \citep{kogut_96a,kogut_96b}, 
OVRO \citep{leitch_97}, Saskatoon \citep{oliveira_97}, 19~GHz \citep{oliveira_98} or Tenerife \citep{oliveira_99}, 
have revealed the presence of a statistical correlation between microwave maps and infrared maps tracing the thermal dust emission.
These observations motivated a search for similar signals in individual regions \citep{finkbeiner_02}, which led to 
unambiguous detections in several molecular clouds with different experiments, like COSMOSOMAS \citep{watson_05}, 
AMI \citep{ami_09a,ami_09b}, CBI \citep{casassus_06,castellanos_11}, VSA \citep{tibbs_10} or Planck \citep{planck_11}.
These studies led to the idea that this dust-correlated signal, which was thenceforth referred to as ``anomalous microwave 
emission'' (AME), was indeed an additional diffuse foreground component, originated by a emission mechanism different from the 
well-known synchrotron, free-free and thermal dust emissions.

AME, which is a significant contaminant of the cosmic microwave background (CMB) in the range $\sim~ 20-60$~GHz, was first 
thought to be free-free emission
from $T_{\rm e}\gtrsim 10^6$~K gas \citep{leitch_97}. This high temperature would be required to reconcile the observed microwave
intensity with the lack of intense H$\alpha$ emission, as it would be expected in the case of bremsstrahlung emission with lower
gas temperatures. However, \citet{draine_98a} ruled out such high temperatures on energetic grounds, and proposed electric 
dipole emission from small rotating dust grains in the interstellar medium (the so-called {\it spinning dust}), a 
mechanism first proposed by \citet{erickson_57}, as a plausible explanation for AME. The models provided by \citet{draine_98b} 
for different media show peaked spectra with maximum emissivities at $\sim 20-50$~GHz which reproduce fairly well the 
observations \citep{finkbeiner_04,oliveira_04,watson_05,iglesias_05,iglesias_06,casassus_06,casassus_08,dickinson_09,tibbs_10}. 
Recently, the previous analytical models have been refined by different groups by introducing more detailed considerations 
about the grain shapes and their rotational properties \citep{ali_09,hoang_10,hoang_11,silsbee_11}.

\citet{draine_99} presented an alternative explanation for AME based on magnetic dipole radiation from hot ferromagnetic 
grains. Models of magnetic dipole emission based on single-domain grains, which predict polarization fractions significantly 
larger than those of the electric dipole emission \citep{lazarian_00}, have been ruled out by different observations 
\citep{battistelli_06,casassus_06,kogut_07,mason_09,lopez_11}.

\citet{bennett_03} proposed an alternative mechanism based on a flat-spectrum synchrotron radiation associated with star 
formation activity to explain Wilkinson Microwave Anisotropy Probe (WMAP) first-year observations. However, this 
hypothesis was questioned by other studies. 
De Oliveira-Costa at al. (2004) argued that the spinning dust hypothesis is clearly 
favored as it can explain the downturn seen in the dust-correlated signal below 20~GHz \citep{fernandez_06,hildebrandt_07}, 
whereas \citet{davies_06} found significant correlation in regions away from star-forming regions. 
More recently \citet{kogut_11} concluded that spinning dust models can fit the ARCADE data at 3, 8 and 10~GHz better than 
the flat-spectrum model.

In this article we present a study of the AME arising in the Pleiades reflection nebula, using data at $\sim 11-17$~GHz from the 
COSMOSOMAS experiment and at $\sim 20-60$~GHz from WMAP. This nearby region, having various bright stars capable of exciting the 
circumstellar medium, is an obvious AME candidate. Its relatively high Galactic latitude ($b = -24\degr$) makes it also 
attractive as we can largely avoid contamination from the Galactic plane. Furthermore, it has been 
widely studied, there being extensive information in the literature about its physical properties that allow us to better understand 
and to constrain the AME intensities. As we will see next, those studies normally refer to different dust substructures 
located within this region. However, the low angular resolution ($\sim 1\degr$) of the 
microwave data we handle preclude us from studying those individual features, and therefore we will be focused in the whole 
complex, to which we will refer to as the ``Pleiades reflection nebula''.


\section{Pleiades enviroment and physical conditions}{\label{sec:pleiades}}

\begin{figure}
\begin{center}
\includegraphics[angle=0,scale=.43]{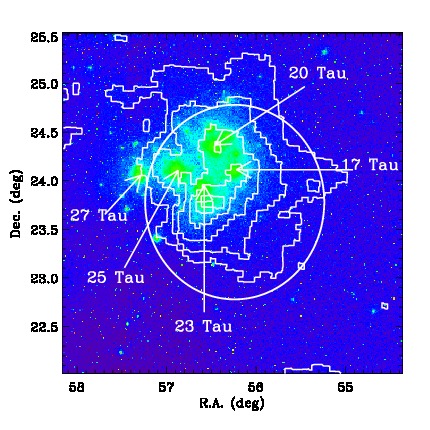}
\caption{Digitized sky survey blue-band photometry\footnote{{\tt http://archive.stsci.edu/cgi-bin/dss\_form}. This image was 
taken with the Oschin Schmidt 48-inch Telescope on Palomar Mountain.} 
in the region of the Pleiades star cluster with IRAS 100~$\mu$m data overlaid with contours. These contours correspond to 
intensity levels of 20, 30, 50, 100, 200 and 300~MJy~sr$^{-1}$. Arrows point to the positions of the main exciting stars in the 
region, and the circle indicates the aperture we will use for flux integration.
}
\label{dss_iras}
\end{center}
\end{figure}

The discovery of a reflection nebulosity towards the Pleiades star cluster \citep{tempel_61} revealed the presence of 
interstellar dust in this region. This dust is reflecting light from hot blue stars in this cluster, especially the bright 
stars 17 Tau, 20 Tau, 23 Tau and 25 Tau, with B-band magnitudes ranging from 2.81 to 4.11 \citep{ritchey_06} and respective 
spectral types B6 III, B7 III, B6 IV and B7 III \citep{white_01}.
The Pleiades cluster 
lies within the Taurus complex at a distance of 125~pc \citep{vanleeuwen_99}, and 
extends $\sim 30'$ ($\sim$1~pc) around coordinates R.A.= 3$^{\rm h}$46$^{\rm m}$, Dec.= 23$\degr$47' (J2000).
\citet{castelaz_87} discovered infrared dust emission around the positions of the Pleiades stars, 
which they attribute to nonequilibrium very small grains or to molecules with 10-100 carbon atoms. The brightness distribution 
of this emission, which extends $\sim 1\degr$, and its connection with the exciting stars, can be appreciated in 
Figure~\ref{dss_iras}, where we show a DSS blue-band image with 100~$\mu$m IRAS intensity contours overplotted. The strong emission 
$\sim 15'$ south of 23 Tau is originated in the Merope molecular cloud, identified by \citet{federman_84} in CO maps. 
Another two bright knots of infrared emission are seen in the IRAS maps concentrated around the stars 17 Tau and 20 Tau.

Except for the area of the Merope molecular cloud, with $E_{\rm B-V}=0.35$~{\rm mag} measured in the direction of the background star 
HD23512 by \citet{white_01}, the optical depth in this region is low ($E_{\rm B-V}\sim 0.03-0.07$~{\rm mag}), and there seem to be little 
other dust along the cluster line of sight \citep{cernis_87}. The physical properties of the gas in different positions of the 
whole complex have been studied by different authors. \citet{federman_84} obtained gas densities $n\sim 300-500$~cm$^{-3}$ and 
temperatures $T_{\rm g}\sim 20$~K from molecular line observations in the Merope molecular cloud. They found column densities 
intermediate between diffuse and dark clouds, and concluded that Merope is indeed much less opaque than typical Taurus clouds. 
\citet{white_84} studied a larger area in the Pleiades enclosing the main exciting stars and estimated densities 
$n\sim 400$~cm$^{-3}$, whereas \citet{gordon_84} found $n\sim 100$~cm$^{-3}$ in a similar area. \citet{ritchey_06} obtained even lower 
values of $n\sim 46$, 16 and 40~cm$^{-3}$ in the lines of sight of 25 Tau, 27 Tau and 28 Tau, respectively.

Different studies, based on CO maps \citep{federman_84} and on high-resolution spectra towards Pleiades stars 
\citep{white_03}, have shown the gas and dust that make up the Pleiades nebula to have a significant radial velocity offset relative 
to the star cluster. A corollary of this is that the Pleiades 
stars were not formed out of this surrounding material but, more likely, the spatial association between the stars and 
the interstellar gas is the result of a chance encounter between the cluster and one or more approaching clouds. 
\citet{white_84} identified two separated phases: one layer of atomic gas with (local standard of rest) radial velocity 
of 7~km~s$^{-1}$, lying $\gtrsim 1$~pc in front of the cluster and with thickness of $\lesssim 0.3$~pc, and one molecular 
phase (though strongly ionized) with radial velocity of 10~km~s$^{-1}$ within the cluster. The presence of these two 
phases {\rm was} confirmed by latter studies, which also showed that the atomic phase probably belongs to the Taurus clouds, 
but are inconclusive on whether or not the molecular phase is embedded within the star cluster \citep{white_03,gibson_03}.

\section{Data analyses}
\subsection{Expected free-free emission from H$\alpha$}
\begin{figure*}
\begin{center}
\includegraphics[angle=0,scale=0.6]{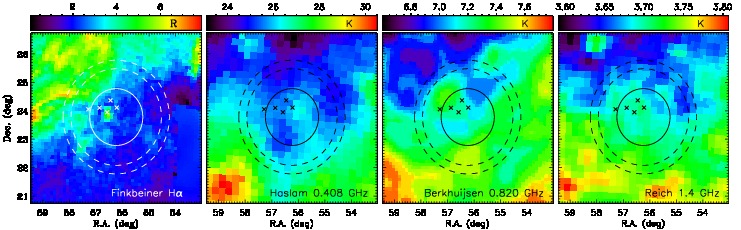}
\caption{H$\alpha$ intensity and radio temperature maps in the position of the Pleiades reflection nebula. 
Crosses mark the positions of the same stars shown in Figure~\ref{dss_iras}. A small knot of H$\alpha$ emission 
is seen $\sim 10'$ south of 23 Tau, but in the radio maps its associated free-free emission is overshadowed by 
the noise and background fluctuations.
}
\label{maps_ff}
\end{center}
\end{figure*}

{\label{sec:ff_est}}
The radio free-free emission must be accurately characterized in order to quantify any possible dust-correlated excess at 
microwave frequencies. To this aim we follow the method described in \citet{dickinson_03}, which allows to estimate the level 
of this emission from H$\alpha$. We resort to the full-sky H$\alpha$ map compiled by \citet{finkbeiner_03}, a patch of 
which is shown in Figure~\ref{maps_ff} in the region of the Pleiades reflection nebula. The solid circle depicts the 1$\degr$ aperture we will use in 
this work for flux integration, while the area limited by the two dashed circles at radii 1.7$\degr$ and 2.0$\degr$ 
define the annulus where we will average the background signal that will be subtracted.
However, in this case it is evident that this annulus would not give a reliable estimation of the background, as  
it contains clear contaminant signal rather different from the true background level around the Pleiades clouds. 
In fact, the averaged signal is larger in the background annulus than in the aperture, giving a negative 
background-subtracted H$\alpha$ intensity of $-0.41\pm 0.09$~R. When positioning the background annulus at different radii 
between 1.0$\degr$ and 3.0$\degr$ we consistently get negative values between $\approx -0.2$~R and $\approx -0.6$~R. Therefore, 
from an aperture integration here we can only set upper limits on the H$\alpha$ emission. Instead, we choose to proceed in a 
different way.
 
Inside the aperture the only evident signal above the background level is a bright H$\alpha$ knot located $\sim 10'$ south of 
23 Tau and coincident with the position of the Merope molecular cloud. On these grounds we consider that all the H$\alpha$ 
emission associated with the Pleiades nebula is produced in this position. In order to calculate its associated intensity 
we fit this feature with a two-dimensional elliptical Gaussian function plus a constant background 
level. We obtain a maximum amplitude (source plus background) of 5.90~R, a background 
level of 2.58~R, and FWHM of 14.3' and 20.7' in two orthogonal directions. Using these parameters the mean 
background-subtracted H$\alpha$ intensity in the 
aperture is $I_{\rm H\alpha}=0.092$~R. Note that this value is higher than the 3-sigma upper limit obtained through 
aperture photometry. Therefore, it is more conservative, as it will be extrapolated to microwave frequencies in order to 
characterize any possible excess emission in that range.

This value needs correction from Galactic extinction, 
$I_{\rm H\alpha}^{\rm corr}=I_{\rm H\alpha}\times 10^{A(\rm H\alpha)/2.5}$. 
Using \citet{odonnell_94} polynomials that allow calculating the extinction at any given wavelength in the range 
$\sim 0.3-0.9~\mu$m, and taking $R_{\rm V}=A_{\rm V}/E_{\rm B-V}=3.6$ from \citet{cernis_87}, we obtain $A$(H$\alpha$)$=2.94~E_{\rm B-V}$. 
Ideally we should apply a pixel by pixel correction, for which we would need a reddening map. We have inspected the 
\citet{schlegel_98} full-sky $E_{\rm B-V}$ map in this region, which gives a mean reddening inside the aperture 
of 0.30~mag, with a maximum of 2.06~mag in a position $\sim 15'$ south of 23 Tau, coincident with the Merope molecular cloud. 
\citet{ritchey_06} give reddening values towards twenty stars belonging to the Pleiades (taken from White et al. 2001), 
with an average value of 
0.074~mag, and a maximum of 0.35~mag towards the star HD23512. These figures are consistent with other measurements 
in the direction of Pleiades or background stars \citep{white_84,cernis_87,herbig_01}. The reddenings from the 
\citet{schlegel_98} map correspond to the full integral along the line of sight, and therefore their higher values 
could be due to the extinction produced in background dust. However, given the relatively high Galactic latitude of the 
Pleiades, one would expect those differences to be smaller. Therefore, instead of performing a pixel by pixel 
correction, we conservatively assume an average value of $E_{\rm B-V}=0.1$, which is slightly larger than the mean over the 
twenty stars of \citet{ritchey_06}. We finally obtain
$I_{\rm H\alpha}^{\rm corr}=0.12$~R, and assuming an electron temperature of 8000~K, we obtain an emission measure of 
$EM= 0.267$~cm$^{-6}$~pc. From this, we derive the expected spectrum of free-free flux integrated over the aperture.

\subsection{Radio data}{\label{sec:radio}}

There are different radio surveys from which we can also constrain the free-free emission. We use the \citet{haslam_82} map 
at 0.408~GHz \footnote{We use the map supplied by \citet{platania_03}.}, the \citet{berkhuijsen_72} map at 0.820~GHz 
\footnote{Downloaded from\\ {\tt http://www.mpifr-bonn.mpg.de/survey.html}} and the \citet{reich_86} map at 
1.4~GHz \footnote{These maps (and, in general, all used in this work) are projected into the HEALP{\sc ix} \citep{gorski_05} pixelization scheme, 
using a pixel size of $\sim7'$ ($N_{\rm side}=512$.). 
In the case of the 0.820~GHz data we carried out the re-pixelization ourselves, as they are given in a different format.}. 
These maps smoothed to a common resolution of 1$\degr$ (except the 0.820~GHz map, whose original resolution is 
1.2$\degr$) in the position of the Pleiades are shown in Figure~\ref{maps_ff}. No obvious excess above the background and noise levels 
is seen, so we have to establish upper limits for the flux densities at these frequencies. To this aim, we assume the noise per 
pixel to be equal to the standard 
deviation of the signal calculated in the background region (external annulus) of the maps smoothed to 1$\degr$. The flux 
noises in the aperture derived from this assumption are respectively 0.38, 0.30 and 0.17~Jy (68\% C.L.; in Table~\ref{fluxes} we 
show upper limits at the 99.7\% C.L.), while the free-free fluxes estimated from H$\alpha$ (see previous section) at those 
frequencies are at a level of $\sim 0.030-0.027$~Jy.


\subsection{Microwave data}
\subsubsection{COSMOSOMAS}{\label{sec:cosmosomas}}
The COSMOSOMAS ({\it Cosmo}logical {\it S}tructures {\it o}n {\it M}edium {\it A}ngular {\it S}cales) experiment consisted of 
two circular-scanning instruments, so-called COSMO11 and COSMO15, operative between 1998 and 2008
from the Teide Observatory (altitude 2400~m, Tenerife). They produced 0.8-1.1$\degr$ resolution daily maps with full 
coverage in right ascension and $\sim$20$\degr$ coverage in declination in four frequency bands centered at 10.9~GHz for 
COSMO11, and 12.7, 14.7 and 16.3~GHz for COSMO15. A description of the instrumental setup of COSMO11 and COSMO15 can be found 
in \citet{hildebrandt_07} and \citet{gallegos_01}, respectively.

An important step of the data processing is the removal of the first seven harmonics
of each scan, in order to suppress the $1/f$ noise from the receiver and from the atmosphere. Each scan, which is a stack of 
30~s of data (corresponding to 30~spins of the primary mirror), is Fourier-fitted and terms up to seventh order (i.e. a 
constant term plus seven sin and cos terms) are removed. This results in a distortion of the broad angular scales and in a 
loss of flux. Further details on the data reduction, calibration and map-making of COSMO11 and COSMO15 data can be found in 
\citet{hildebrandt_07} and in \citet{fernandez_06}, respectively. In those references the final maps can be seen, as well as 
the discussion about the dust-correlated signal that was found at $|b|>20\degr$.

The maps we use here are a stack of $\sim 150$~days of observations in each of the two linear polarization modes of COSMO11 
(November 2003 to June 2005) and of $\sim 110$~days of COSMO15 data (October 1999 to January 2000).
They are convolved to a common resolution of 1.12$\degr$, and have final sensitivities (averaged over the 
$\approx$10,000 square-degree observed sky region) of 35~$\mu$K~beam$^{-1}$ for COSMO11, and 53, 56 and 
118~$\mu$K~beam$^{-1}$ for the three frequencies of COSMO15.

COSMO11 and COSMO15 observed in different configurations, covering different declination ranges. The minimum declination 
was $\approx 23.5\degr$. In the case of COSMO15, there are also $\sim$5 days in a configuration down to 
Dec.=16.7$\degr$, but we discard these data because of being too noisy. Most of the area encompassed by the Pleiades 
reflection nebula is covered by the observations, but unfortunately its center lies just 17' from the border of the map. 
Also, the total number of days of observations in the configurations reaching these low declinations 
are $\sim 95$ for COSMO11 and $\sim 12$ for COSMO15. 
However, despite the fewer observing days, the maps sensitivities in the region of the Pleiades are similar to the 
full-map average sensitivities quoted above thanks to the circular scanning strategy, which results in a better spatial 
coverage towards the edge of the map. Possible edge effects arisen in this region are avoided by ignoring the 
blank pixels in the convolution process.

\begin{figure*}
\begin{center}
\includegraphics[angle=0,scale=0.6]{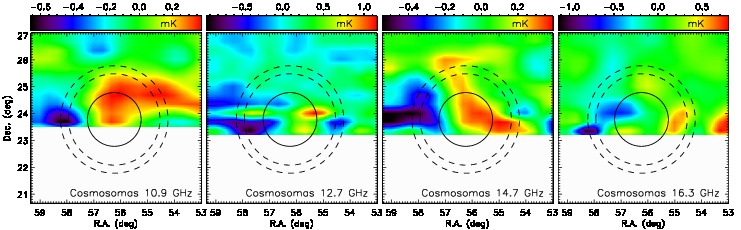}
\caption{COSMOSOMAS maps towards the Pleiades reflection nebula, for the four frequencies of this experiment. 
Note that here we are displaying data down to a declination of $\approx23.5\degr$. There are data in a different 
configuration which extends the declination coverage to 16.7$\degr$, but with $\sim 20$ times less integration time, 
and therefore with a notably higher noise level. 
}
\label{maps_cosmo}
\end{center}
\end{figure*}

COSMOSOMAS maps in the region of the Pleiades are shown in Figure~\ref{maps_cosmo}. No clear detection is seen in any of 
the frequency bands. Marginal evidence for a weak signal in the position of the Pleiades could be claimed in the COSMO11 
map. We obtain an indirect estimate of $0.67\pm 0.09$~Jy for the 11~GHz flux from the DIRBE fluxes calculated in 
section~\ref{sec:fir}, by using the dust-correlated emissivities between the COSMO11 map and the DIRBE 100, 140 and 
240~$\mu$m maps. However, we conservatively establish flux upper limits for COSMO11 
and for the three COSMO15 frequencies, which will in turn be important to trace the AME flux downturn at 
lower frequencies predicted by spinning-dust models. To this aim we measure the standard deviation in each 
map within a box of $\approx 10$~square~degrees around the Pleiades, obtaining 65, 96, 65 and 72~$\mu$K~beam$^{-1}$ 
for 10.9, 12.7, 14.7 and 16.3~GHz, respectively. After transforming to flux, these values are corrected 
for the flux suppression caused by the COSMOSOMAS map-making strategy described above. To this aim, we simulate a 
source with the same brightness distribution as the 100~$\mu$m DIRBE map in the position of the Pleiades, and a flux 
normalized to 1~Jy. This source is then degraded to an angular resolution of 1.12$\degr$, and convolved with the simulated 
circular scanning strategy of COSMOSOMAS, which includes removal of the first seven harmonics. By measuring the flux 
in this final map, and comparing with the initial flux of 1~Jy, we conclude that the flux losses are 62\% 
and 66\% respectively for COSMO11 and COSMO15. Final upper limits are listed in Table~\ref{fluxes} at the 99.7\% 
confidence level (C.L.).

\subsubsection{WMAP intensity}
\begin{figure}
\begin{center}
\includegraphics[angle=0,scale=0.43]{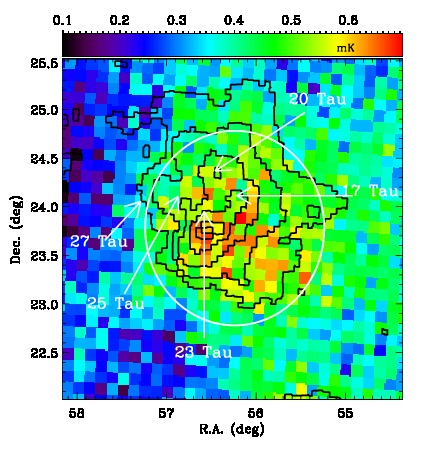}
\caption{WMAP K-band (22.8~GHz) map with IRAS 100$\mu$m data overplotted with contours, in the position of the Pleiades nebula. 
Arrows indicate the positions of the brightest stars in the region, and the circle the 1$\degr$-radius aperture we use for 
flux integration. Temperature units of the WMAP map are indicated in the 
color scale in the top. IRAS contours correspond to intensity levels of 20, 30, 50, 100, 200 and 300~MJy~sr$^{-1}$. 
A clear correlation is seen between the microwave (AME) and the infrared (thermal dust emission) intensities.}
\label{wmap_iras}
\end{center}
\end{figure}

\begin{figure*}
\begin{center}
\includegraphics[angle=0,scale=0.6]{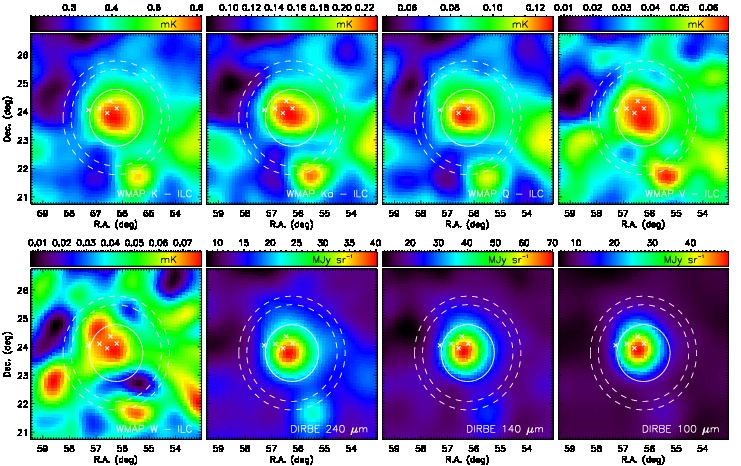}
\caption{WMAP and DIRBE maps smoothed to a common resolution of 1$\degr$. WMAP maps have been subtracted from the CMB 
contribution given by the ILC template (see text for details). 
Crosses mark the positions of the same stars shown in Figure~\ref{dss_iras}. The 1$\degr$-radius 
circular aperture and 1.7-2.0$\degr$ circular annulus which are used for flux extraction are indicated. 
Strong AME is evident in WMAP maps, with maximum intensity at 22.8~GHz.}
\label{wmap_dirbe}
\end{center}
\end{figure*}

{\label{sec:wmap_int}}
The NASA's WMAP satellite have produced full-sky maps at frequency bands centered at 22.8, 33.0, 40.7, 60.8 and 93.5~GHz 
(K, Ka, Q, V and W), and with angular resolutions of 0.85$\degr$, 0.65$\degr$, 0.51$\degr$, 0.35$\degr$ and 0.25$\degr$, 
respectively. We use the seven-year data products \citep{jarosik_11}, available in the {\sc lambda} 
database\footnote{\tt http://lambda.gsfc.nasa.gov/}. In Figure~\ref{wmap_iras} we {\rm represent} a composition of the K-band map at 
the position of the Pleiades, with IRAS 100~$\mu$m contours overplotted, showing an evident spatial correlation between both 
maps. The Pearson correlation coefficient calculated between maps at the same angular resolution (1$\degr$) is $r=0.608$. 
However, the center of the microwave emission seems to be shifted $\sim 15'$ to the southwest of the position of the 
stars 23 Tau and 17 Tau, around which the bulk of the the infrared emission is seen. 
Figure~\ref{wmap_dirbe} shows WMAP smoothed maps subtracted from the CMB contribution (see next paragraph). AME shows 
up prominently especially in the K band. The bulk of the emission, which extends $\sim 1.5\degr$, seem to be mounted on a 
larger-scale diffuse structure which elongates towards the southwest of the maps. 
Some emission is also hinted at in the  0.408-1.4~GHz maps (see Figure~\ref{maps_ff}). We calculate spectral indices  
between 0.408 and 22.8~GHz in different circular apertures on this diffuse region surrounding the Pleiades nebula, obtaining 
values around $-0.5$, which are compatible with synchrotron emission. However, we can not draw a firm conclusion about the possible 
contribution from AME owing to the low flux density and to the contamination from the stronger emission arisen within the 
Pleiades dust. Radial profiles of the WMAP K-band map at its original angular resolution
across the source in different directions show the 22.8~GHz emission to extend $\sim 1\degr$, being slightly resolved by 
the 0.85$\degr$-FWHM beam. This further supports the hypothesis that, even if the bulk of the microwave emission may be coming 
from a position $\sim 15'$ southwest of 23 Tau and 17 Tau, there may be an important diffuse component originated in a wider 
region encompassing this nebulosity. 

\begin{figure}
\begin{center}
\includegraphics[angle=0,scale=0.6]{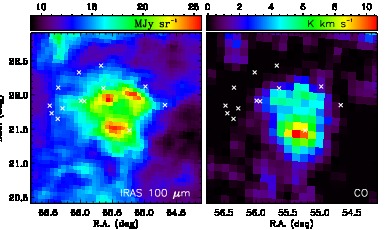}
\caption{IRAS $100~\mu$m and CO maps of the source found towards the southwest of the Pleiades, with coordinates 
(R.A.,Dec.)=$55.44\degr +21.67\degr$. Crosses mark the positions of the infrared sources from the IRAS catalogue.
}
\label{sw_source}
\end{center}
\end{figure}

Towards the southwest of the Pleiades stars, at equatorial coordinates 
(R.A.,Dec.)=$55.44\degr +21.67\degr$, WMAP maps also reveal a 
source, with a flux of $\approx 0.6$~Jy at 22.8~GHz, which has no counterpart in the 1.4~GHz NVSS \citep{condon_98} nor in the 
WMAP point source catalogues \citep{gold_11}. It does not have significant emission in the $0.408-1.4$~GHz maps nor in the 
H$\alpha$ map of \citet{finkbeiner_03}. 
The flux upper limits 
that can be derived from the radio data, and the flux value at 22.8 GHz, are consistent with a free-free spectrum.
WMAP 93.5~GHz data show evidence of thermal dust emission.
Moreover, as it is shown in Figure~\ref{sw_source}, this source is visible 
in IRAS 60 and 100~$\mu$m maps as a circular ring with a diameter of $\approx 40$', as well as in the CO map of \citet{dame_01}. 
There are several infrared sources around that position listed in the IRAS point source catalogue. The spectrum of WMAP 
and DIRBE data is compatible with thermal dust emission. 

For the sake of consistency, we perform the flux extraction in maps at the same angular resolution. We then use
{\sc lambda} maps degraded to a common resolution of 1$\degr$. Onwards we will consider two separate strategies for flux 
calculation: i) case A: we apply our method for flux extraction directly on those maps, which may contain some CMB 
contamination; ii) case B: we construct CMB-subtracted maps on which our method for flux extraction will be applied.
To define the CMB template that will be pixel-by-pixel subtracted from the original maps, rather than using the full-sky 
internal linear combination (ILC) map supplied by the WMAP team, we generate our own ILC in a circular area of 
30$\degr$-radius around the Pleiades after masking out point sources by applying 
WMAP's KQ85 mask. We use the same technique as the WMAP team, i.e. a weighted linear combination of the five smoothed frequency maps, in which 
the weights are chosen to minimize the variance of the measured temperatures, in order to maintain the CMB contribution 
while minimizing the foreground contribution. Pixel temperatures in our ILC are given by 
$T_{\rm ILC}$ = $-0.1937~T_{\rm K}$ $-0.0550~T_{\rm Ka}$ $+0.8289~T_{\rm Q}$ $+0.7042~T_{\rm V}$ $-0.28437~T_{\rm W}$, 
and the resulting CMB-subtracted maps are shown in Figure~\ref{wmap_dirbe}.

In order to extract the fluxes we use a direct aperture integration. This is a well-known 
technique \citep{bennett_93,hernandez_04,planck_11}, consisting on integrating temperatures of all pixels within a given 
aperture, and subtracting a background level which is calculated by averaging pixel temperatures enclosed by an external 
annulus. The flux is then given by
\begin{equation}
S_\nu=a(\nu)\left[\frac{\sum_{i=1}^{n1}T_{i}}{n_{1}}-\frac{\sum_{j=1}^{n2}T_{j}}{n_{2}}\right]~~,
\end{equation}
where $n_1$ and $n_2$ are respectively the number of pixels in the circular aperture and in the external annulus, and $T_i$ and 
$T_j$ represent the pixel thermodynamic temperatures in the aperture and in the background annulus. The function $a(\nu)$ gives the 
conversion factor from temperature to flux,
\begin{equation}
a(\nu) = \frac{h^2\nu^4}{2 k_{\rm b} T_{\rm cmb}^2 c^2}~{\rm sinh}^{-2}\left(\frac{h\nu}{2 k_{\rm b}T_{\rm cmb}}\right)n_1\Omega_{\rm pix}~~,
\end{equation}
where $h$ and $k_{\rm b}$ are the Planck and Boltzmann constants, 
$T_{\rm cmb}=2.725$~K \citep{mather_99} the CMB temperature and $\Omega_{\rm pix}$ the solid angle subtended by each pixel. 
The associated error 
bar is calculated through the quadratic sum of the variance of each pixel (given by WMAP error maps) plus the pixel-to-pixel 
covariance:
\[
\sigma(S_\nu) = a(\nu)\left[\frac{1}{n_1^2}\left(\sum_{ij}^{n1 n1}C(\theta_{ij})+\sum_i^{n1}\sigma_i^2\right)+\right.
\]
\begin{equation}
\left.\frac{1}{n_2^2}\left(\sum_{ij}^{n2 n2}C(\theta_{ij})+\sum_{i}^{n2}\sigma_{i}^2\right)-
\frac{2}{n_1 n_2}\sum_{ij}^{n1 n2}C(\theta_{ij})\right]^{1/2}~~.
\end{equation}
In this equation $\sigma_i$ represents the instrumental noise associated to each pixel.
$C(\theta_{ij})$ is the noise correlation function, which arises from the smoothing process, and is evaluated 
for any pair of pixels separated an angle $\theta_{ij}$. For case A, we also calculate the error associated with the CMB
primordial fluctuations, by introducing its 
correlation function, derived from a theoretical power spectrum obtained from the WMAP best-fit cosmological model.

The radii we use for the aperture and for the internal and external circles defining the annulus are shown in Figure~\ref{wmap_dirbe}.
These are respectively 1.0$\degr$, 1.7$\degr$ and 2.0$\degr$, and were chosen from the radial angular profiles of the signal. 
These angular profiles flatten at the distance of the background annulus, and the aperture encloses the bulk of the microwave and 
infrared emissions associated with the Pleiades. 
The center coordinates of the aperture are R.A.= 3$^{\rm h}$44$^{\rm m}$58$^{\rm s}$, Dec.= 23$\degr$46'48'' (J2000). 
This position is chosen to match the maximum emission at 22.8~GHz, and is located $\approx 15'$ southwest of the middle 
of the line connecting the 17~Tau and 23~Tau stars, around which the bulk of the infrared emission is located. 
Final fluxes are shown in Table~\ref{fluxes}. Error bars include only the thermal noise, and the contribution from the CMB 
variance is indicated within brackets for case A. In case B the flux error bar includes the ILC noise, which is calculated in each 
map pixel through the quadratic sum of pixel noises for all bands weighted by the corresponding ILC coefficient. 
The flux at 22.8~GHz calculated in the ILC map, which is subtracted in case B, is $0.45\pm 0.11$~Jy. This value is 
compatible with 0.51~Jy, which corresponds to the standard deviation of the CMB primordial fluctuations associated with 
this measurement. 

At 22.8~GHz we get $2.15\pm 0.12$~Jy for case B, which is 
two orders of magnitude above the level of free-free emission at this frequency derived from H$\alpha$ (0.02~Jy; see 
section~\ref{sec:ff_est}). We will see later in section~\ref{sec:sed} that this flux is in fact dominated by AME, as 
the contribution from thermal dust emission is rather low at this frequency. Also, considering the upper limit from COSMO11, the 
spectral index between 10.9 and 22.8~GHz is $\beta_{11-23}\gtrsim 1.12$, a value consistent with typical spinning dust
models.

\subsubsection{WMAP polarization}
Estimates or constraints of the AME polarization can help to discriminate between models based on electric dipole emission 
from very small spinning dust grains and magnetic dipole radiation from hot ferromagnetic grains 
(see section~\ref{sec:intro} and references therein). It is therefore interesting to get estimates of the polarization 
from WMAP maps of Q and U Stokes parameters. Applying the same aperture photometry technique as in section~\ref{sec:wmap_int} in 
these maps, which show no significant emission at the position of the Pleiades, we get
fluxes of $Q=0.076\pm 0.071$~Jy and $U=-0.056\pm 0.091$~Jy at 22.8~Ghz. Using these values, together with the flux density 
for case B (see previous section), we derive a 95\% C.L. upper limit on the polarization fraction, 
$\Pi=\sqrt{Q^2+U^2}/I$, of $\Pi<10.8\%$. This constraint is much weaker than $\Pi<1.0\%$ 
obtained in the much brighter region of the Perseus molecular cloud \citep{lopez_11}.

\subsection{Far-infrared data}{\label{sec:fir}}
Far-infrared data trace the thermal dust emission, which can then be extrapolated to microwave frequencies. We use 
Zodi-Subtracted Mission Average (ZSMA) COBE-DIRBE maps \citep{hauser_98} at 240~$\mu$m (1249~GHz), 140~$\mu$m (2141~GHz) 
and 100~$\mu$m (2998~GHz). Although there are DIRBE data at higher frequencies, we only consider these three bands, because 
they allow modelling the thermal dust spectrum with a single modified black-body curve. Higher-frequency bands are dominated 
by a different grain population and their inclusion would make necessary at least two curves. 
We also note that, in spite of IRAS finer angular resolution, we use here DIRBE data because IRAS 
lower-frequency band is 100~$\mu$m, and also because we calculate fluxes in a larger angular scale than either of IRAS and DIRBE 
beam sizes.

In Figure~\ref{wmap_dirbe} we show the DIRBE maps at the position of the Pleiades smoothed to a resolution of 
1$\degr$. It is again clear from these plots that the microwave emission is slightly shifted to the southwest with respect to 
the infrared emission at 100~$\mu$m. This offset is less pronounced at 140~$\mu$m, and almost disappears at 240~$\mu$m. 
The Pearson correlation coefficients between WMAP 22.8~GHz and DIRBE bands from 100~$\mu$m to 240~$\mu$m are respectively 
$r=0.585$, 0.722 and 0.821. However, our aperture encloses the bulk of the flux in both frequency ranges.
Aperture-photometry fluxes in the three bands are quoted in Table~\ref{fluxes}. We also show the dust-correlated 
emissivities, which are used to quantify the correlation between the 100~$\mu$m DIRBE map and WMAP maps. These are calculated 
by a linear fit of the temperature ($\mu$K) of WMAP background-subtracted pixels to the intensity (MJy~sr$^{-1}$) of 
DIRBE background-subtracted pixels using a standard least-squares method. A significant correlation is found, specially 
between WMAP 22.8~GHz and DIRBE 100~$\mu$m channels, where we get $4.36\pm 0.17$~$\mu$K/(MJy~sr$^{-1}$). We note however 
that this dust emissivity is strikingly lower than those obtained in dust clouds \citep{davies_06}, which usually range from $\sim 11$ 
to $\sim 35$~$\mu$K/(MJy~sr$^{-1}$), and more similar to the $3.3\pm 1.7$~$\mu$K/(MJy~sr$^{-1}$)  
found by \citet{dickinson_07} in HII regions. \citet{vidal_11} got an even lower value of 
$0.2\pm 0.1$~$\mu$K/(MJy~sr$^{-1}$) in the translucent cloud LDN~1780, where they found evidence of AME. 

\begin{table*}
\begin{center}
\scriptsize{
\caption{Fluxes and dust-correlated emissivities}
\label{fluxes}
\begin{tabular}{ccccccccc}
\hline\hline
\noalign{\smallskip}
 && \multicolumn{3}{c}{Case A} & & \multicolumn{3}{c}{Case B}\\
\cline{3-5}\cline{7-9}
\noalign{\smallskip}
$\nu$ && Flux & Residual flux & Correlation && Flux & Residual flux & Correlation \\
(GHz) && (Jy)  & (Jy)  & $\mu$K/(MJy~sr$^{-1}$) && (Jy)  & (Jy)  & $\mu$K/(MJy~sr$^{-1}$)  \\
\noalign{\smallskip}
\hline
\noalign{\smallskip}
0.408  && $<$ 1.14  & $<$ 1.11  & -  &&  $<$ 1.14  & $<$ 1.11  & - \\     
0.820  && $<$ 0.89  & $<$ 0.87  & -  &&  $<$ 0.89  & $<$ 0.87  & - \\  
1.42   && $<$ 0.51  & $<$ 0.49  & -  &&  $<$ 0.51  & $<$ 0.49  & - \\  
10.9   && $<$ 1.04  & $<$ 0.87  & -  &&  $<$ 0.94  & $<$ 0.91  & - \\
12.7   && $<$ 1.97  & $<$ 1.75  & -  &&  $<$ 1.83  & $<$ 1.80  & - \\
14.7   && $<$ 1.77  & $<$ 1.48  & -  &&  $<$ 1.58  & $<$ 1.56  & - \\
16.3   && $<$ 2.43  & $<$ 2.08  & -  &&  $<$ 2.20  & $<$ 2.17  & - \\
22.8   &&  2.60$\pm$0.06 ($\pm$ 0.51) &  1.95$\pm$0.06   &  3.01$\pm$0.27 && 2.15 $\pm$ 0.12  & 2.12 $\pm$ 0.12 & 4.36 $\pm$ 0.17  \\
33.0   &&  2.55$\pm$0.10 ($\pm$ 1.06) &  1.21$\pm$0.12   &  0.66$\pm$0.17 && 1.61 $\pm$ 0.15  & 1.55 $\pm$ 0.15 & 2.01 $\pm$ 0.09  \\
40.7   &&  2.64$\pm$0.15 ($\pm$ 1.59) &  0.64$\pm$0.17  & -0.32$\pm$0.16 && 1.24 $\pm$ 0.18  & 1.12 $\pm$ 0.18 & 1.03 $\pm$ 0.03  \\
60.8   &&  4.71$\pm$0.36 ($\pm$ 3.37) &  0.39$\pm$0.40  & -0.77$\pm$0.16 && 1.75 $\pm$ 0.38  & 1.23 $\pm$ 0.38 & 0.59 $\pm$ 0.02  \\
93.5   &&  9.12$\pm$0.89 ($\pm$ 7.03) & -0.52$\pm$0.97  & -0.25$\pm$0.12 && 2.94 $\pm$ 0.90  & 0.37 $\pm$ 0.90 & 1.10 $\pm$ 0.05  \\
1249.1 && 11931$\pm$185  &  9$\pm$394 &    --  &&11931 $\pm$ 185 & -366 $\pm$ 195  & -- \\
2141.4 && 23469$\pm$249  & -14$\pm$595 &    --  &&23469 $\pm$ 249 &  618 $\pm$ 262  & -- \\
2997.9 && 17959$\pm$ 89  &  1$\pm$375 &    --  &&17959 $\pm$  89 &  -47 $\pm$  101  & --  \\
\noalign{\smallskip}
\hline
\noalign{\smallskip}
\end{tabular}
}
\end{center}
{\bf Notes}. Flux upper limits at the 99.7\% C.L. level are shown for the 0.408-1.42~GHz and COSMOSOMAS (10.9-16.3~GHz) data. 
For WMAP and DIRBE frequencies fluxes are obtained through an aperture integration, using a radius 
of 1$\degr$ for the aperture and subtracting a mean background calculated in an annulus between 1.7$\degr$ and 2.0$\degr$ around the source.
The fluxes on the right (case B) were calculated on maps where a pixel-by-pixel subtraction of a CMB internal linear combination map
was applied. The error bars include instrumental noise, and the numbers within brackets show errors associated with the CMB anisotropies. 
The residual fluxes represent the level of AME, as they were obtained after the subtraction of the rest of the modelled components. 
We also show the dust-correlated emissivities, which indicate the correlation between WMAP and the IRAS 100~$\mu$m map.\\
\end{table*}

\section{Spectral energy distribution}
\begin{figure}
\includegraphics[angle=0,scale=.45]{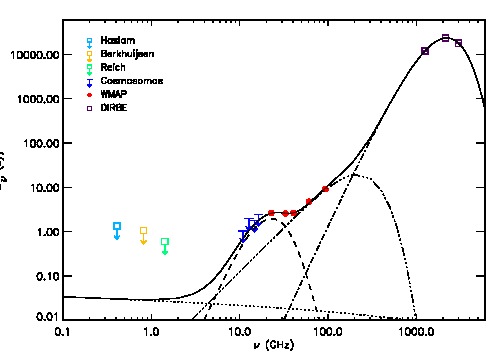}
\includegraphics[angle=0,scale=.45]{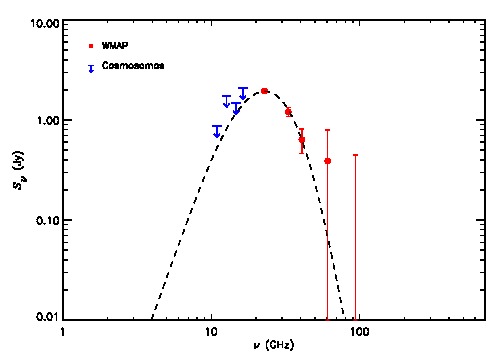}
\caption{Top: spectral energy distribution (SED) of the Pleiades nebula for case A. Upper limits are shown at the 99.7\% C.L. 
for radio and COSMOSOMAS data. WMAP and DIRBE fluxes have been obtained through aperture photometry, and are quoted 
in Table~\ref{fluxes}. The fitted model consists of free-free emission (dotted line), thermal dust (dash-dotted line), 
CMB (dash-triple-dotted line) and a spinning dust model for a molecular gas phase (dashed line). The model 
parameters are listed in Table~\ref{params}. Bottom: residual SED after subtraction of the free-free 
and the best-fit CMB and thermal dust components.}
\label{sed_cmb}
\end{figure}

\begin{figure}
\includegraphics[angle=0,scale=.45]{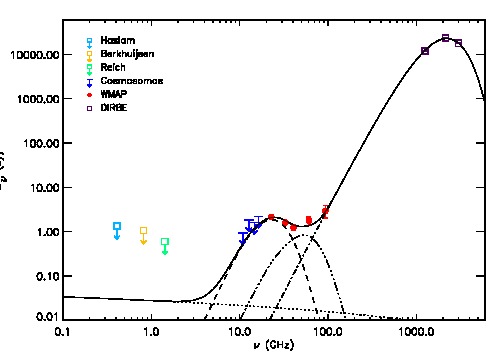}
\includegraphics[angle=0,scale=.45]{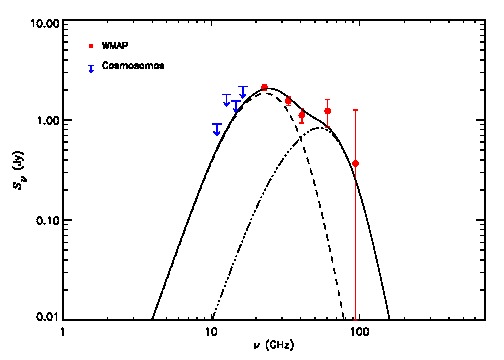}
\caption{Top: same as in Figure~\ref{sed_cmb}, but for case B, i.e. a pixel-by-pixel subtraction of the CMB is performed, and 
then no CMB component is considered in the SED fitting. Fluxes for this case are quoted in Table~\ref{fluxes}. The 
fitted model consists of free-free emission (dotted line), thermal dust (dash-dotted line), and two spinning dust components: high-density molecular gas (dashed line) 
and low-density atomic gas (dashed-triple-dotted line). The model parameters are listed in Table~\ref{params}. 
Bottom: residual SED after subtraction of the free-free and best-fit thermal dust components.}
\label{sed_nocmb}
\end{figure}

{\label{sec:sed}}
Figures~\ref{sed_cmb} and \ref{sed_nocmb} depict the spectral energy distributions (SED) for cases A and B, respectively.
In both cases, we represent 3-sigma (99.7\% C.L.) upper limits at 0.408, 0.820 and 1.42~GHz, as no significant emission was found at these 
frequencies (see section~\ref{sec:radio}). Upper limits at the 99.7\% C.L. are also shown for COSMO11 and for the three channels of 
COSMO15 (see section~\ref{sec:cosmosomas}). 

\begin{table}
\begin{center}
\caption{Model parameters}
\label{parameters}
\label{params}
\scriptsize{
\begin{tabular}{lcc}
\hline\hline
\noalign{\smallskip}
$T_{\rm e}$ (K) & \multicolumn{2}{c}{8000} \\
$EM$ (cm$^{-6}$pc) & \multicolumn{2}{c}{0.267} \\
\noalign{\smallskip}
\hline
\noalign{\smallskip}
& Molecular & Atomic \\
\noalign{\smallskip}
\hline
\noalign{\smallskip}
$n_{\rm H}$ (cm$^{-3}$)  & 300  & 200  \\ 
$T_{\rm g}$ (K)      & 20   & 1000 \\
$\chi$               & 0.03 & 10   \\ 
$x_{\rm H}$ (ppm)      & 9.2  & 373  \\
$x_{\rm C}$ (ppm)      & 1    & 100  \\
$y$                    & 1    & 0.1  \\
$\beta$ (D)          & 9.34 & 9.34 \\
\noalign{\smallskip}
\hline
\noalign{\smallskip}
& \multicolumn{2}{c}{Case A} \\
\hline
\noalign{\smallskip}
$N_{\rm H}$ (10$^{20}$ cm$^{-2}$) & 6.94 $\pm$ 0.22 & --  \\
$\tau_{100}$ & \multicolumn{2}{c}{(6.09 $\pm$ 0.06)$\times$10$^{-4}$}  \\
$\beta_{\rm d}$ & \multicolumn{2}{c}{2.29 $\pm$ 0.02}  \\
$T_{\rm d}$ (K) & \multicolumn{2}{c}{20.12 $\pm$ 0.03}\\
$\Delta T_{\rm cmb}$ ($\mu$K) & \multicolumn{2}{c}{42.2 $\pm$ 1.9}\\
\noalign{\smallskip}
\hline
\noalign{\smallskip}
& \multicolumn{2}{c}{Case B} \\
\hline
\noalign{\smallskip}
N$_{\rm H}$ (10$^{20}$ cm$^{-2}$) & 6.60 $\pm$ 0.11 & 0.30 $\pm$ 0.01  \\
$\tau_{100}$ & \multicolumn{2}{c}{(3.302 $\pm$ 0.004)$\times$10$^{-4}$}  \\
$\beta_{\rm d}$ & \multicolumn{2}{c}{1.869 $\pm$ 0.004}  \\
$T_{\rm d}$ (K) & \multicolumn{2}{c}{22.008 $\pm$ 0.005}\\
\noalign{\smallskip}
\hline
\noalign{\smallskip}
\end{tabular}
}
\end{center}
{\bf Notes.} The two values on the top of the table define the spectrum of the free-free emission. $T_{\rm e}$ is an assumed value, and $EM$ is 
calculated from the H$\alpha$ intensity corrected from Galactic extinction (see section~\ref{sec:ff_est}). 
Parameters from $n_{\rm H}$ to $\beta$ (see section~\ref{sec:sed} for an explanation of their meaning) are used 
as inputs for the spinning-dust models. They are fixed at typical values for the two 
different gas phases (molecular and atomic). The 
rest of the parameters are jointly fitted to the spectrum. $N_{\rm H}$ defines the amplitude of the spinning dust component; 
$\tau_{100}$, $\beta_{\rm d}$ and $T_{\rm d}$ define the thermal dust spectrum; and $\Delta T_{\rm cmb}$ is the average value of the 
CMB temperature within the aperture. This CMB component is fitted only in case A, as in case B we calculate the fluxes on 
CMB-subtracted maps. Also, in case A no atomic gas phase producing spinning-dust emission was considered.
\end{table}

AME, which neatly shows up in the SEDs as a bump at frequencies $\sim$20-60~GHz, is usually well fitted by spinning-dust 
models (see section~\ref{sec:intro} and references therein). Following \citet{planck_11}, we consider two complementary 
spinning-dust components, corresponding respectively to high-density molecular gas and to low-density atomic gas.
Different studies of the dynamics of the Pleiades environment agree on the existence of these two gas phases, the 
former being probably located within the cluster and the later lying $\sim 1$~pc in front of it 
(see section~\ref{sec:pleiades} and references therein). We compute spinning-dust emissivities per hydrogen column 
density of these two phases using 
the {\sc spdust.2} code\footnote{http://www.tapir.caltech.edu/$\sim$yacine/spdust/spdust.html} \citep{ali_09,silsbee_11}. 
Rather than fitting all the spinning dust parameters to the observed spectra, we use realistic values and fit the amplitude 
to the spectrum. Assumed physical parameters for these two phases, that are used as inputs for {\sc spdust.2}, are shown 
in Table~\ref{parameters}. $n_{\rm H}$ is the total hydrogen number density, $T_{\rm g}$ the kinetic gas temperature, $\chi$ 
is the intensity of the 
radiation field relative to the average interstellar radiation field, $x_{\rm H}=n_{\rm H}^+/n_{\rm H}$ represents the 
hydrogen ionization fraction, $x_{\rm C}=n_{\rm C}^+/n_{\rm H}$ the ionized carbon fractional abundance, 
$y=2n({\rm H_2})/n_{\rm H}$ the molecular hydrogen fractional abundance and $\beta$ the average dipole moment per atom. 
Parameters for the grain-size distribution are taken from line 25 of Table~1 of \citet{weintgartner_01}.

Different studies of the Pleiades interstellar material cited in section~\ref{sec:pleiades} provide 
temperatures around $T_{\rm g}\sim 20$~K and densities spanning from $n\sim 40$~cm$^{-3}$ \citep{ritchey_06} to 
$\sim 400$~cm$^{-3}$ \citep{white_84}. This allows flexibility when fixing our model parameters. Moreover, we note that some of 
the aforementioned studies focused on individual clouds whereas here we study the emission integrated over an area encompassing 
the whole Pleiades reflection nebula. The physical parameters we choose for the molecular phase are similar to those 
of the molecular cloud environment of \citet{draine_98b}, while those of the atomic phase have been adapted to have a 
spectrum peaking at a higher frequency, in order to provide a better fit to the 60.8~GHz point. The position of the 
peak of the spinning dust spectrum is very sensitive to the intensity of the radiation field. For the molecular phase, which 
is probably associated with the Merope molecular cloud, we select a low value of $\chi =0.03$, in order to have the 
peak at $\approx 23$~GHz, matching the WMAP K-band flux, which is our maximum measured value. 

\citet{ritchey_06} have calculated the intensity of the radiation field in different 
lines of sight towards the Pleiades; from their values (see their Table~6) in positions within our aperture we 
calculate an average of $\chi\approx 16$. Using this value and a hydrogen density of $n_{\rm H} = 30.6$~cm$^{-3}$ (also 
calculated from an average of their measurements), and same values for the rest of the {\sc spdust.2} input parameters, 
we obtain a spinning dust spectrum which peaks at 36.9~GHz and provides a very poor fit to the data. If we use for the 
hydrogen number density a value as low as $n_{\rm H} = 0.03$~cm$^{-3}$, while maintaining the same radiation field intensity, the 
spectrum peaks at a lower value of 31~GHz, still far from the observed peak around 22.8~GHz. Therefore, in order 
to reliably fit the data, we have to decrease the intensity of the radiation field for the molecular phase. Note however that 
\citet{ritchey_06} do not 
provide measurements for the radiation field in any position towards the Merope molecular cloud, and therefore it could be 
the case that its intensity is indeed lower. On the other hand, for the atomic phase, 
which probably traces a more diffuse component scattered around the clouds, we adopt 
$\chi=10$, which is closer to the values reported by \citet{ritchey_06}, and also provides a better fit to the 60.8~GHz point. 
Once we have obtained the spinning dust spectra for the molecular and atomic phases from {\sc spdust.2}, their amplitudes, which 
are defined by the hydrogen column densities $N_{\rm H}$, are fitted to the measured fluxes.

We model the thermal dust, which is the main contribution at infrared frequencies, through a modified black-body 
spectrum, $I^{\rm d}_\nu (T_{\rm d})=\tau_{100}~(\nu/(2998~{\rm GHz}))^{\beta_{\rm d}}~B_\nu(T_{\rm d})$, with $T_{\rm d}$ being the dust 
temperature, $\tau_{100}$ the optical depth at 100~$\mu$m and $\beta_{\rm d}$ the emissivity spectral index. The 
free-free spectrum is normalized by assuming an electron temperature $T_{\rm e}=8000$~K, and estimating the electron measure, 
indicated in Table~\ref{parameters}, through the extinction-corrected H$\alpha$ emission in the region 
(see section~\ref{sec:ff_est}). This spectrum is represented by the dotted lines of Figures~\ref{sed_cmb} and \ref{sed_nocmb}.

In the SED of case A (Figure~\ref{sed_cmb}) the CMB is probably contaminating our measurements and boosting the WMAP V and W-band 
fluxes. Therefore, in this case we also fit to the data a CMB component with an average temperature $\Delta T_{\rm cmb}$, and 
neglect the spinning-dust corresponding to the atomic gas phase, as it would be completely overshadowed by the CMB component. 
After subtracting the free-free flux corresponding to each frequency, we jointly fit, using a non-linear least-squares method, 
our five model parameters, $N_{\rm H}$, $\Delta T_{\rm cmb}$, $T_{\rm d}$, $\tau_{100}$ and $\beta_{\rm d}$, defining the 
spinning-dust, CMB and thermal dust components, to the WMAP and DIRBE fluxes. In case B (Figure~\ref{sed_nocmb}), we fit the 
hydrogen column density associated with the spinning-dust atomic phase instead of $\Delta T_{\rm cmb}$, as the CMB has been 
removed directly from the map. 
The fitted parameters and their errors are shown in Table~\ref{parameters}, and the spectra of all 
components are represented in Figures~\ref{sed_cmb} and \ref{sed_nocmb}. We see that these fitted spectra are indeed fully 
compatible with the COSMOSOMAS 3-sigma upper limits. The reduced chi-squares are 0.75 and 6.76 for cases A and B, respectively. 

The dust temperature is somewhat higher in case B, with the spectrum being steeper in case A. This is a 
consequence of $\Delta T_{\rm cmb}$ in case B being slightly larger than the average ILC temperature within the aperture, 
30.2~$\mu$K, which is the effective CMB level that we are subtracting in case A prior to flux calculation.
The fitted values for the hydrogen column densities are compatible in both cases. 
These values show that the Pleiades reflection nebula constitute indeed a more diffuse media than others were AME have been 
usually studied, like the Perseus and $\rho$ Ophiuchus molecular clouds, where $N_{\rm H}= 117\times 10^{20}$~cm$^{-2}$ and 
$171\times 10^{20}$~cm$^{-2}$ respectively \citep{planck_11}. This also becomes evident from the reddening measurements in this 
region, $E_{\rm B-V}=0.1$~{\rm mag} (see section~\ref{sec:ff_est}), which are typical of diffuse clouds, and lower than the 
respective values in the Perseus and $\rho$ Ophiuchus molecular clouds, and even lower than the reddening of the translucent 
cloud LDN1780, $E_{\rm B-V}\sim 0.6$, where \citet{vidal_11} recently found AME. Our fitted values for the hydrogen 
column densities are also of the same order of those derived by applying the scaling relation of 
\citet{bohlin_78}, ($N_{\rm H}+N_{\rm H2})/E_{\rm B-V}=5.8\times 10^{21}$~cm$^{-2}$~mag$^{-1}$, and taking for the reddening 
$E_{\rm B-V}=0.1$~{\rm mag}. It may also be noted that by dividing the fitted hydrogen column density of the molecular phase by 
the hydrogen number density, we get an estimate for the depth of the cloud of $\sim 0.7$~pc, a value that agrees with the characteristic 
length scale of this cloud.

It is also interesting to compare our results with those presented in Figure~13 of  
\citet{vidal_11}, where they plot emissivity at 31~GHz (intensity at 31 GHz divided by hydrogen column density) versus hydrogen 
column density. Dividing the total flux at 33~GHz for case B by the solid angle of the aperture and by the 
corresponding fitted column density we get 
$\epsilon_{33}=(2.55\pm 0.28)\times 10^{-24}$~MJy~sr$^{-1}$~cm$^{2}$. In order to scale to 31~GHz we use our fitted model, 
and obtain $\epsilon_{31}=(3.03\pm 0.33)\times 10^{-24}$~MJy~sr$^{-1}$~cm$^{2}$. This value roughly follows the trend shown in 
that figure, but falls below the fitted line, which for $N_{\rm H}=6.60\times 10^{20}$~cm$^{-2}$ predicts 
$\sim 5.7\times 10^{-24}$~MJy~sr$^{-1}$~cm$^{2}$. It therefore seems that the Pleiades nebula, in addition to having a notable 
lower hydrogen column density to the regions shown in that figure, and actually to the majority of the regions where AME has been 
previously detected, also has a lower emissivity per hydrogen nucleon.

The residual fluxes, obtained after subtracting all modelled components except
the spinning-dust are quoted in Table~\ref{fluxes} and plotted in bottom panels of Figures~\ref{sed_cmb} and 
\ref{sed_nocmb}. 
According to our fitted model for case B, the thermal dust contributes at 22.8~GHz with a flux of 
0.012~Jy, whereas the level of the free-free at this frequency was 0.019~Jy. Then, the residual flux at this frequency, 
which can be attributed to AME, is 2.12$\pm$0.12~Jy (17.7$\sigma$; modelling errors included).

\section{Conclusions}
We have presented evidence for anomalous microwave emission in the position of the nearby (125~pc) Pleiades star cluster, one 
of the most studied regions in the sky. This AME is originated in the dust grains that make up the Pleiades reflection 
nebula, whose thermal infrared emission was discovered by \citet{castelaz_87}. Using data from the seven-year release of WMAP, 
here we have measured a total flux density towards this nebulosity of $2.15\pm 0.12$~Jy at 22.8~GHz by applying an aperture 
integration in a 1$\degr$-radius circle around the position (R.A.,Dec.)=$56.24\degr +23.78 \degr$ (J2000). 
COSMOSOMAS maps at 11-17~GHz do not show evidence for emission at this position. Nevertheless, we have set upper 
limits (99.7\% C.L.) of 0.94~Jy at 10.9~GHz and 1.58~Jy at 13.6~GHz, which help to trace the downturn of the AME 
spectrum at these frequencies predicted by models based on electric dipole emission from fast-spinning dust grains. 

In order to quantify the total AME flux we have considered other emission mechanisms. We have estimated a free-free flux of 0.019~Jy, at the 
same reference frequency of 22.8~GHz, from the H$\alpha$ map 
of \citet{schlegel_98}, which has been previously corrected from Galactic extinction by using reddening values 
measured towards different stars. Infrared fluxes obtained from the COBE-DIRBE experiment have been used to trace the thermal dust 
emission, from which we extrapolate to microwave frequencies and obtain a flux of 0.012~Jy. Therefore, both the free-free and 
thermal dust emissions have little impact in our measurement. After deducting these two components we have obtained a residual flux of 
$2.12\pm 0.12$~Jy, which is a detection of AME at the 17.7$\sigma$ level. 

The signal at 22.8~GHz is slightly resolved by WMAP K-band beam (0.85$\degr$ FWHM). 
This indicates that, although the bulk of the emission is probably coming from the region near the stars 23 Tau and 17 Tau, 
where the strongest infrared emission is also detected, there is probably an important diffuse component integrated over the 
whole nebulosity. Observations with microwave experiments at higher angular resolution are crucial to elucidate 
the relative contribution to the total observed flux of the different individual clouds that constitute the Pleiades reflection nebula. 
They may also help to better trace the correlation with the thermal dust, and to shed light on the properties of the grain 
population that may be originating this AME.

After deducting the extrapolated free-free fluxes, we have fitted a joint model consisting of a modified blackbody spectrum, 
tracing the thermal dust emission, and a spinning-dust spectrum which traces the AME, to the WMAP and DIRBE fluxes. 
For the spinning-dust component we have considered two complementary phases, one dominated by high-density molecular gas and the 
other by lower-density atomic gas. Their spinning-dust spectra have been obtained 
with {\sc spdust.2}, by using realistic values for the physical parameters that are inputs to this code. We then fitted 
the amplitude of the spectra to the data, together with the three parameters that define the thermal dust model. The 
resulting spinning-dust fit successfully traces the WMAP and DIRBE data, while being compatible with the COSMOSOMAS upper limits.

Microwave temperatures at 22.8~GHz are clearly correlated with 100~$\mu$m infrared intensities, the dust emissivity being 
$4.36\pm 0.17$~$\mu$K/(MJy~sr$^{-1}$). This value is lower than what is typical in cool dust clouds, and more 
characteristic of HII regions. The physical properties of the Pleiades nebula show that this is a much less opaque medium 
than others where AME has been generally studied, the reason for this detection likely being its proximity. For instance, 
the optical depth and the hydrogen column density of the Perseus molecular cloud are more than an order of magnitude higher 
\citep{watson_05,planck_11}, which explains why the AME flux at 22.8~GHz is there twenty times larger than in the Pleiades 
despite being at twice their distance. This potentially makes the Pleiades an attractive environment to study the AME emission 
in different physical conditions than those generally explored up to now. 


\acknowledgments

We thank the referee for useful comments, which have helped to extend 
the discussion on some important aspects. 
The color-scale image of Figure~\ref{dss_iras} is 
based on photographic data obtained using Oschin Schmidt Telescope      
on Palomar Mountain. The Palomar Observatory Sky Survey was funded     
by the National Geographic Society. The Oschin Shmidt Telescope is     
operated by the California Institute of Technology and Palomar           
Observatory.  The plates were processed into the present compressed     
digital format with their permission. The Digitized Sky Survey was     
produced at the Space Telescope Science Institute (ST ScI) under        
U. S. Government grant NAG W-2166. We acknowledge the use of the MPIfR Survey Sampler website at {\tt http://www.mpifr-bonn.mpg.de/survey.html}, from where we got 
the 0.820~GHz data. 
We acknowledge the use of the Legacy Archive for Microwave Background Data Analysis (LAMBDA). Support for LAMBDA is 
provided by the NASA Office of Space Science. Some of the results in this paper have been obtained using the 
HEALP{\sc ix} \citep{gorski_05} package. This work has been partially funded by project AYA2010-21766-C03-02 of the
Spanish Ministry of Science and Innovation (MICINN).
JAR-M is a Ram\'on y Cajal fellow of the MICINN. \\



\clearpage


\begin{thebibliography}{}

\bibitem[Ali-Ha{\"i}moud et al.(2009)]{ali_09} 
Ali-Ha{\"i}moud, Y., Hirata, C.~M., 
\& Dickinson, C.\ 2009, \mnras, 395, 1055 

\bibitem[Ami Consortium: Scaife et al.(2009a)]{ami_09a} Ami Consortium: Scaife, A.~M.~M., 
Hurley-Walker, N., Green, D.~A., et al.\ 2009a, \mnras, 394, L46 

\bibitem[Ami Consortium: Scaife et al.(2009b)]{ami_09b} Ami Consortium: Scaife, A.~M.~M., 
Hurley-Walker, N., Green, D.~A., et al.\ 2009, \mnras, 400, 1394 


\bibitem[Battistelli et al.(2006)]{battistelli_06} Battistelli, E.~S., 
Rebolo, R., Rubi{\~n}o-Mart{\'{\i}}n, et al.\ 2006, \apjl, 645, L141 

\bibitem[Bennett et al.(1993)]{bennett_93} Bennett, C.~L., 
Hinshaw, G., Banday, A., et al.\ 1993, \apjl, 414, L77 

\bibitem[Bennett et al.(2003)]{bennett_03} Bennett, C.~L., Hill, R.~S., Hinshaw, G., et al.\ 
2003, \apjs, 148, 97 

\bibitem[Berkhuijsen(1972)]{berkhuijsen_72} Berkhuijsen, E.~M.\ 1972, \aaps, 5, 263

\bibitem[Bohlin et al.(1978)]{bohlin_78} Bohlin, R.~C., Savage, 
B.~D., \& Drake, J.~F.\ 1978, \apj, 224, 132 

\bibitem[Casassus et al.(2006)]{casassus_06} Casassus, S., Cabrera, 
G.~F., F{\"o}rster, et al.\ 2006, \apj, 639, 951 

\bibitem[Casassus et al.(2008)]{casassus_08} Casassus, S., Dickinson, C., Cleary, K.,
et al.\ 2008, \mnras, 391, 1075 

\bibitem[Castelaz et al.(1987)]{castelaz_87} Castelaz, M.~W., 
Sellgren, K., \& Werner, M.~W.\ 1987, \apj, 313, 853 

\bibitem[Castellanos et al.(2011)]{castellanos_11} Castellanos, P., Casassus, S., Dickinson, S., et 
al.\ 2011, \mnras, 411, 1137 

\bibitem[Cernis(1987)]{cernis_87} Cernis, K.\ 1987, \apss, 133, 355 

\bibitem[Condon et al.(1998)]{condon_98} Condon, J.~J., Cotton, 
W.~D., Greisen, E.~W., Yin, et al.\ 1998, \aj, 115, 1693 

\bibitem[Dame et al.(2001)]{dame_01} Dame, T.~M., Hartmann, D., 
\& Thaddeus, P.\ 2001, \apj, 547, 792 

\bibitem[Davies et al.(2006)]{davies_06} Davies, R.~D., 
Dickinson, C., Banday, et al.\ 2006, \mnras, 370, 1125 

\bibitem[de Oliveira-Costa et al.(1997)]{oliveira_97} de 
Oliveira-Costa, A., Kogut, A., Devlin, M.~J., et al.\ 1997, \apjl, 482, L17 

\bibitem[de Oliveira-Costa et al.(1998)]{oliveira_98} de 
Oliveira-Costa, A., Tegmark, M., Page, L.~A., 
\& Boughn, S.~P.\ 1998, \apjl, 509, L9 

\bibitem[de Oliveira-Costa et al.(1999)]{oliveira_99} de 
Oliveira-Costa, A., Tegmark, M., Gutierrez, C.~M., et al.\ 1999, \apjl, 527, L9 

\bibitem[de Oliveira-Costa et al.(2004)]{oliveira_04} de 
Oliveira-Costa, A., Tegmark, M., Davies, R.~D., et al.\ 2004, \apjl, 606, L89 

\bibitem[Dickinson et al.(2003)]{dickinson_03} Dickinson, C., 
Davies, R.~D., \& Davis, R.~J.\ 2003, \mnras, 341, 369 


\bibitem[Dickinson et al.(2007)]{dickinson_07} Dickinson, C., 
Davies, R.~D., Bronfman, L., et al.\ 2007, \mnras, 379, 297 

\bibitem[Dickinson et al.(2009)]{dickinson_09} Dickinson, C., Davies, R.~D., Allison, J.~R., et 
al.\ 2009, \apj, 690, 1585 

\bibitem[Dickinson et al.(2010)]{dickinson_10} Dickinson, C., Casassus, S., Davies, R.~D., et 
al.\ 2010, \mnras, 407, 2223 

\bibitem[Draine 
\& Lazarian(1998a)]{draine_98a} Draine, B.~T., \& Lazarian, A.\ 1998a, \apjl, 494, L19 

\bibitem[Draine 
\& Lazarian(1998b)]{draine_98b} Draine, B.~T., \& Lazarian, A.\ 1998b, \apj, 508, 157 

\bibitem[Draine 
\& Lazarian(1999)]{draine_99} Draine, B.~T., \& Lazarian, A.\ 1999, \apj, 512, 740 

\bibitem[Erickson(1957)]{erickson_57} Erickson, W.~C.\ 1957, \apj, 
126, 480 

\bibitem[Federman 
\& Willson(1984)]{federman_84} Federman, S.~R., \& Willson, R.~F.\ 1984, \apj, 283, 626 

\bibitem[Fern{\'a}ndez-Cerezo et al.(2006)]{fernandez_06} 
Fern{\'a}ndez-Cerezo, S., Guti\'errez, C.~M., Rebolo, R., et al.\ 2006, \mnras, 370, 15 

\bibitem[Finkbeiner et al.(2002)]{finkbeiner_02} Finkbeiner, D.~P., 
Schlegel, D.~J., Frank, C., \& Heiles, C.\ 2002, \apj, 566, 898 

\bibitem[Finkbeiner(2003)]{finkbeiner_03} Finkbeiner, D.~P.\ 2003, 
\apjs, 146, 407 

\bibitem[Finkbeiner(2004)]{finkbeiner_04} Finkbeiner, D.~P.\ 2004, 
\apj, 614, 186 

\bibitem[Gallegos et al.(2001)]{gallegos_01} Gallegos, J.~E., 
Mac{\'{\i}}as-P{\'e}rez, J.~F., Guti{\'e}rrez, C.~M., et al.\ 2001, \mnras, 327, 1178 

\bibitem[Gibson 
\& Nordsieck(2003)]{gibson_03} Gibson, S.~J., \& Nordsieck, K.~H.\ 2003, \apj, 589, 362 

\bibitem[Gold et al.(2011)]{gold_11} Gold, B., Odegard, N., Weiland, J.~L., et al.\ 2011, 
\apjs, 192, 15 

\bibitem[Gordon 
\& Arny(1984)]{gordon_84} Gordon, K.~J., \& Arny, T.~T.\ 1984, \aj, 89, 672 

\bibitem[G{\'o}rski et al.(2005)]{gorski_05} G{\'o}rski, K.~M., 
Hivon, E., Banday, A.~J., et al.\ 2005, \apj, 622, 759 

\bibitem[Haslam et al.(1982)]{haslam_82} Haslam, C.~G.~T., Salter, C.~J., Stoffel, H., \& Wilson, W.~E.\ 1982, \aaps, 47, 1

\bibitem[Hauser et al.(1998)]{hauser_98} Hauser, M.~G., Arendt, R.~G., Kelsall, T., et al.\ 
1998, \apj, 508, 25 

\bibitem[Herbig 
\& Simon(2001)]{herbig_01} Herbig, G.~H., \& Simon, T.\ 2001, \aj, 121, 3138 

\bibitem[Hern{\'a}ndez-Monteagudo 
\& Rubi{\~n}o-Mart{\'{\i}}n(2004)]{hernandez_04} Hern{\'a}ndez-Monteagudo, C., \& Rubi{\~n}o-Mart{\'{\i}}n, J.~A.\ 2004, \mnras, 347, 403 

\bibitem[Hildebrandt et al.(2007)]{hildebrandt_07} Hildebrandt, S.~R., 
Rebolo, R., Rubi{\~n}o-Mart{\'{\i}}n, J.~A., et al.\ 2007, \mnras, 382, 594 

\bibitem[Hoang et al.(2010)]{hoang_10} Hoang, T., Draine, B.~T., 
\& Lazarian, A.\ 2010, \apj, 715, 1462 

\bibitem[Hoang et al.(2011)]{hoang_11} Hoang, T., Lazarian, A., 
\& Draine, B.~T.\ 2011, arXiv:1105.2302 

\bibitem[Iglesias-Groth(2005)]{iglesias_05} Iglesias-Groth, S.\ 
2005, \apjl, 632, L25 

\bibitem[Iglesias-Groth(2006)]{iglesias_06} Iglesias-Groth, S.\ 
2006, \mnras, 368, 1925 


\bibitem[Jarosik et al.(2011)]{jarosik_11} Jarosik, N., Bennett, C.~L., Dunkley, J., et al.\ 
2011, \apjs, 192, 14 

\bibitem[Kogut et al.(1996a)]{kogut_96a} Kogut, A., Banday, A.~J., 
Bennett, C.~L., et al.\ 1996a, \apj, 460, 1 

\bibitem[Kogut et al.(1996b)]{kogut_96b} Kogut, A., Banday, A.~J., 
Bennett, C.~L., et al.\ 1996b, \apjl, 464, L5 

\bibitem[Kogut et al.(2007)]{kogut_07} Kogut, A., Dunkley, J., Bennett, C.~L., et al.\ 2007, 
\apj, 665, 355 

\bibitem[Kogut et al.(2011)]{kogut_11} Kogut, A., Fixsen, D.~J., Levin, S.~M., et al.\ 2011, 
\apj, 734, 4 

\bibitem[Lazarian 
\& Draine(2000)]{lazarian_00} Lazarian, A., \& Draine, B.~T.\ 2000, \apjl, 536, L15 

\bibitem[Leitch et al.(1997)]{leitch_97} Leitch, E.~M., Readhead, 
A.~C.~S., Pearson, T.~J., \& Myers, S.~T.\ 1997, \apjl, 486, L23 

\bibitem[L{\'o}pez-Caraballo et al.(2011)]{lopez_11} 
L{\'o}pez-Caraballo, C.~H., Rubi{\~n}o-Mart{\'{\i}}n, J.~A., Rebolo, R., 
\& G{\'e}nova-Santos, R.\ 2011, \apj, 729, 25 

\bibitem[Mason et al.(2009)]{mason_09} Mason, B.~S., Robishaw, 
T., Heiles, C., Finkbeiner, D., \& Dickinson, C.\ 2009, \apj, 697, 1187 

\bibitem[Mather et al.(1999)]{mather_99} Mather, J.~C., Fixsen, 
D.~J., Shafer, R.~A., Mosier, C., \& Wilkinson, D.~T.\ 1999, \apj, 512, 511 


\bibitem[O'Donnell(1994)]{odonnell_94} O'Donnell, J.~E.\ 1994, 
\apj, 422, 158 

\bibitem[Planck Collaboration et al.(2011)]{planck_11} Planck 
Collaboration: Ade, P.~A. R., Aghanim, N., Arnaud, M., et al.\ 2011, arXiv:1101.2031 

\bibitem[Platania et 
al.(2003)]{platania_03} Platania, P., Burigana, C., Maino, D., et al.\ 2003, \aap, 410, 847 

\bibitem[Reich \& Reich(1986)]{reich_86} Reich, P., \& Reich, W.\ 1986, \aaps, 63, 205

\bibitem[Ritchey et al.(2006)]{ritchey_06} Ritchey, A.~M., 
Martinez, M., Pan, K., Federman, S.~R., 
\& Lambert, D.~L.\ 2006, \apj, 649, 788 

\bibitem[Schlegel et al.(1998)]{schlegel_98} Schlegel, D.~J., 
Finkbeiner, D.~P., \& Davis, M.\ 1998, \apj, 500, 525 

\bibitem[Silsbee et al.(2011)]{silsbee_11} Silsbee, K., 
Ali-Ha{\"i}moud, Y., \& Hirata, C.~M.\ 2011, \mnras, 411, 2750 

\bibitem[Tempel(1861)]{tempel_61} Tempel, W.\ 1861, Astronomische 
Nachrichten, 54, 285 

\bibitem[Tibbs et al.(2010)]{tibbs_10} Tibbs, C.~T., Watson, R.~A., Dickinson, C., et al.\ 
2010, \mnras, 402, 1969 

\bibitem[van 
Leeuwen(1999)]{vanleeuwen_99} van Leeuwen, F.\ 1999, \aap, 341, L71 

\bibitem[Vidal et al.(2011)]{vidal_11} Vidal, M., Casassus, S., Dickinson, C., et al.\ 2011, 
\mnras, 414, 2424 

\bibitem[Watson et al.(2005)]{watson_05} Watson, R.~A., Rebolo, 
R., Rubi{\~n}o-Mart{\'{\i}}n, et al.\ 2005, \apjl, 624, L89 

\bibitem[Weingartner 
\& Draine(2001)]{weintgartner_01} Weingartner, J.~C., \& Draine, B.~T.\ 2001, \apj, 548, 296 

\bibitem[White(1984)]{white_84} White, R.~E.\ 1984, \apj, 284, 
685 

\bibitem[White et al.(2001)]{white_01} White, R.~E., Allen, 
C.~L., Forrester, W.~B., Gonnella, A.~M., 
\& Young, K.~L.\ 2001, \apjs, 132, 253 

\bibitem[White(2003)]{white_03} White, R.~E.\ 2003, \apjs, 148, 
487 

\end{thebibliography}
\end{document}